\documentclass[journal]{IEEEtran}
\usepackage{graphicx}
\usepackage[hyphens]{url}
\usepackage{amsmath}
\usepackage{tikz}
\usepackage{pgfplots}
\usetikzlibrary{shadows.blur}
\usetikzlibrary{shapes.symbols}
\usepackage{amsfonts}

\usepackage{xcolor}

\graphicspath{{figs/}}

\begin{document}

\title{Detecting Community Depression Dynamics Due to COVID-19 Pandemic in Australia}

\author{Jianlong Zhou$^*$,~\IEEEmembership{Senior Member,~IEEE,}
         Hamad Zogan,
         Shuiqiao Yang,
         Shoaib Jameel, 
         Guandong Xu$^*$, 
         Fang Chen
        % <-this % stops a space
\thanks{J.~Zhou, S. Yang, and F. Chen are with Data Science Institute, University of Technology Sydney, Australia, e-mails: Jianlong.Zhou@uts.edu.au, Shuiqiao.Yang@uts.edu.au, Fang.Chen@uts.edu.au}% 
\thanks{H.~Zogan and G.~Xu are with Advanced Analytics Institute, University of Technology Sydney, Australia, e-mails: Hamad.A.Zogan@student.uts.edu.au, Guandong.Xu@uts.edu.au}% 
\thanks{S.~Jameel is with University of Essex, UK, email: Shoaib.Jameel@essex.ac.uk}% using \url for the break of the email address into two lines for better presentation.
\thanks{$^*$Corresponding authors}%
%\thanks{Manuscript received June 30, 2020; revised xxxx, 2020.}
}

%\markboth{IEEE Transactions on Computational Social Systems}%


\maketitle

\begin{abstract}
The recent COVID-19 pandemic has caused unprecedented impact across the globe. We have also witnessed millions of people with increased mental health issues, such as depression, stress, worry, fear, disgust, sadness, and anxiety, which have become one of the major public health concerns during this severe health crisis. For instance, depression is one of the most common mental health issues according to the findings made by the World Health Organisation (WHO). Depression can cause serious emotional, behavioural and physical health problems with significant consequences, both personal and social costs included. This paper studies community depression dynamics due to COVID-19 pandemic through user-generated content on Twitter. A new approach based on multi-modal features from tweets and Term Frequency-Inverse Document Frequency (TF-IDF) is proposed to build depression classification models. Multi-modal features capture depression cues from emotion, topic and domain-specific perspectives. We study the problem using recently scraped tweets from Twitter users emanating from the state of New South Wales in Australia. Our novel classification model is capable of extracting depression polarities which may be affected by COVID-19 and related events during the COVID-19 period. The results found that people became more depressed after the outbreak of COVID-19. The measures implemented by the government such as the state lockdown also increased depression levels. Further analysis in the Local Government Area (LGA) level found that the community depression level was different across different LGAs. In an LGA, the severe health emergencies but not the confirmed cases of COVID-19 in the LGA may make people more depressed. Such granular level analysis of depression dynamics not only can help authorities such as governmental departments to take corresponding actions more objectively in specific regions if necessary but also allows users to perceive the dynamics of depression over the time to learn the effectiveness of measures implemented by the government or negative effects of any big events for emergency management.

%We propose a new approach based on term frequency--inverse document frequency to extract multi-modality features from user tweets, which are used to build classification models capable of capturing depression among the users. 
\end{abstract}


% \fi
\IEEEpeerreviewmaketitle

\section{Introduction}

%\tempnotes{Jianlong mainly works on this section.}
%\tempnotes{Since the deadline is coming, I suggest to directly edit the text when doing the proofreading to save time. Shoaib: Yes, that's the right path to choose, many thanks!}

\IEEEPARstart{T}{he} outbreak of the novel Coronavirus Infectious Disease 2019 (COVID-19) has caused an unprecedented impact on people's daily lives around the world \cite{noauthor_coronavirus_2020}. People's lives are at risk because the virus can easily spread from person to person \cite{surveillances2020epidemiological} either by coming in close contact with the infected person or sometimes may even spread through community transmission\footnote{https://www.who.int/publications/i/item/preparing-for-large-scale-community-transmission-of-covid-19}, which then becomes extremely challenging to contain. The infection has now rapidly spread across the world and there have been more than 10.3 million confirmed cases and more than 505,000 people have died because of the infection until 30 June 2020\footnote{https://coronavirus.jhu.edu/}. Almost every country in the world is battling against COVID-19 to prevent it from spreading as much as possible. While some countries such as New Zealand has been very successful in containing the spread, others such as Brazil and India have not. As a result, this outbreak has caused immense distress among individuals either through infection or through increased mental health issues, such as depression, stress, worry, fear, disgust, sadness, anxiety (a fear for one's health, and a fear of infecting others), and perceived stigmatisation \cite{montemurro_emotional_2020, bhat_sentiment_2020, rogers_psychiatric_2020}. These mental health issues can even occur in people, not at high risk of getting infected. There could be even several people who are exposed to the virus may be unfamiliar with it as they may not follow the news, or are completely disconnected with the general population. \cite{montemurro_emotional_2020}.

\begin{figure}[!tb]
  \centering
  \includegraphics[width=0.99\linewidth]{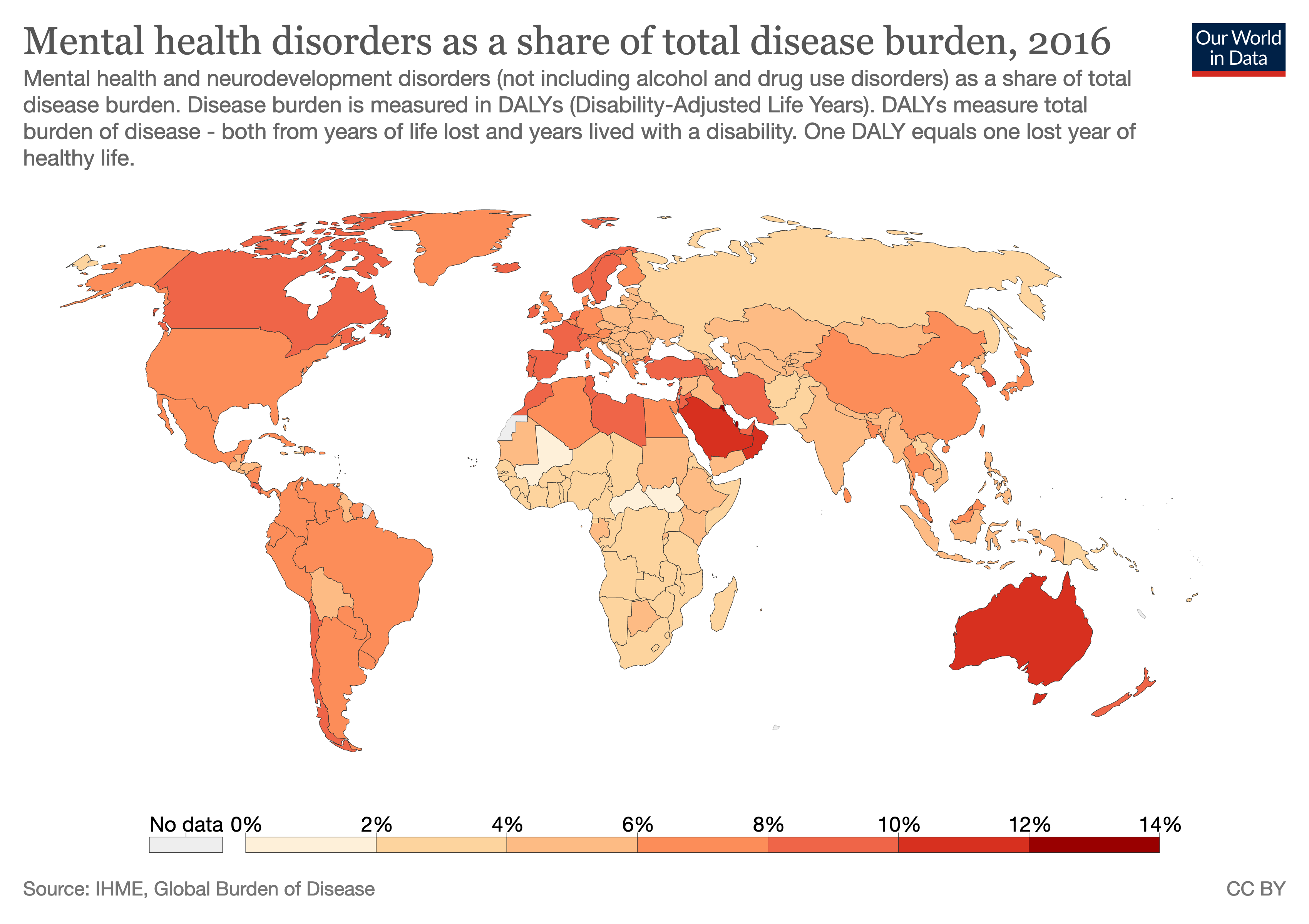}
  \caption{The world mental health disorders in 2016\protect\footnotemark.}
\label{fig:world_mental_health2016}       
\end{figure}

Consider depression as an example, which is the most common mental health issue among other mental health issues according to the World Health Organisation (WHO), with more than 264 million people suffering from the depression worldwide \cite{who_depression_2020}. Australia is one of the top countries where mental health disorders have high proportions over the total disease burden (see Fig.~\ref{fig:world_mental_health2016}). Depression can cause severe emotional, behavioural and physical health problems. For example, people with depression may experience symptoms amounting to their inability to focus on anything, they constantly go through the feeling of guilt and irritation, they suffer from low self-worth, and experience sleep problems. Depression can, therefore, cause serious consequences, at both personal and social costs \cite{vigo_estimating_2016}.

A person can experience several complications as a result of depression. Complications linked to mental health especially depression include: unhappiness and decreased enjoyment of life, family conflicts, relationship difficulties, social isolation, problems with tobacco, alcohol and other drugs, self-harm and harm to others (such as suicide or homicide), weakened immune system. Furthermore, the consequence of depression goes beyond functioning and quality of life and extends to somatic health. Depression has been shown to subsequently increase the risk of, for example, cardiovascular, stroke, diabetes and obesity morbidity \cite{penninx_understanding_2013}. However, the past epidemics can suggest some cues of what to look out for after COVID-19 in the next few months and years. For example, when patients with SARS and MERS were assessed a few months later, 14.9\% had depression and 14.8\% had an anxiety disorder \cite{rogers_psychiatric_2020}. 

\footnotetext{\url{https://ourworldindata.org/}}
% Preliminary data suggest that patients with COVID-19 might experience mental disorders such as delirium, confusion, agitation, and altered consciousness, as well as symptoms of depression, anxiety, and insomnia \cite{rogers_psychiatric_2020}.

% The impact of the COVID-19 outbreak has been so huge that it is said to be the most serious epidemic in the past hundred years comparable to pandemics of the past like Spanish flu of 1918, or the Black Death in the mid-1300s.

Meanwhile, to reduce the risk of the virus spreading among people and communities, different countries have taken strict measures such as locking down the whole city and practising rigorous social-distancing among people. For example, countries such as China, Italy, Spain, and Australia are fighting the COVID-19 pandemic through nation-wide lockdown or by cordoning off the areas that were suspected of having risks of community spread throughout the pandemic, expecting to ``flatten the curve''. However, the long-term social activity restriction policies adopted during the pandemic period may further amplify the mental health issues of people. Therefore, it is important to examine people's mental health states as a result of COVID-19 and related policies, which can help governments and related agencies to take appropriate measures more objectively if necessary.

% further work
On the other hand, we have witnessed increased usage of online social media such as Twitter during the lockdown\footnote{https://www.statista.com/statistics/1106498/home-media-consumption-coronavirus-worldwide-by-country/}. For instance, 40\% consumers have spent longer than usual on messaging services and social media during the lockdown. It is mainly because people are eager to publicly express their feelings online given an unprecedented time that they are going through both physically and emotionally. The social media platforms represent a relatively real-time large-scale snapshot of activities, thoughts, and feelings of people's daily lives and thereby reflect their emotional well-being. Every tweet represents a signal of the users' state of mind and state of being at that moment \cite{gibbons_twitter-based_2019}. Aggregation of such digital traces may make it possible to monitor health behaviours at a large-scale, which has become a new growing area of interest in public health and health care research \cite{cavazos-rehg_content_2016,jaidka_estimating_2020}.

Since social media is social by its nature, and social patterns can be consequently found in Twitter feeds, for instance, thereby revealing key aspects of mental and emotional disorders \cite{coppersmith_quantifying_2014}. As a result, Twitter recently has been increasingly used as a viable approach to detect mental disorders of depression in different regions of the world \cite{reece_forecasting_2017,almouzini_detecting_2019,leis_detecting_2019,razak_tweep_2020,mcclellan2017using}. For example, the research found that the depressed users were less active in posting tweets, doing it more frequently between 23:00 and 6:00. The use of vocabularies could also be an indicator of depression in Twitter, for example, it was found that the use of verbs was more common by depressed users, and the first-person singular pronoun was by far the most used by depressed users \cite{leis_detecting_2019}. Hence, many research work has been done to extract features like user's social activity behaviours, user profiles, texts from their social media posts for depression detection using machine learning techniques \cite{de2013predicting,tsugawa2015recognizing,yang2015gis,orabi_deep_2018,shen2017depression}. For example, De Choudhury \cite{de2013predicting} et al., proposed to predict depression for social media users based on Twitter using support vector machine (SVM) for prediction based on manually labelled training data. 
%\blue{It would be nice to elaborate more on these works here so that a reviewer immediately understands the vital differences and similarities. This paragraph would help win or lose because it is the most crucial paragraph of the paper. For instance, we could mention the datasets, how the analysis was done, any new computational model was proposed. This is the place where we could highlight some key difference between them and us.} 

The most recent work using Twitter to analyse mental health issues due to COVID-19 are \cite{barkur_sentiment_2020,bhat_sentiment_2020,zhou_examination_2020}. These work focus more on public sentiment analysis. Furthermore, little work such as Li et al., \cite{li_what_2020} classify each tweet into the emotions of anger, anticipation, disgust, fear, joy, sadness, surprise and trust. The two emotions of sadness and fear are more related to severe negative sentiments like depression due to COVID-19. However, little work is done to detect depression dynamics at the state level or even more granular level such as suburb level. Such granular level analysis of depression dynamics not only can help authorities such as governmental departments to take corresponding actions more objectively in specific regions if necessary but also allows users to perceive the dynamics of depression over the time to learn the effectiveness of policies implemented by the government or negative effects of any big events. The answers to the questions that we wish to find are:
%Our research questions are \blue{Shall we brand these as, the answers to the questions that we wish to find than research questions?}:
\begin{itemize}
    \item How people's depression is affected by COVID-19 in the time dimension in the state level?
    \item How people's depression is affected by COVID-19 in the time dimension in local government areas?
    \item Can we detect the effects of policies/measures implemented by the government during the pandemic on depression?
    \item Can we detect the effects of big events on depression during the pandemic?
    \item How effective is the model in detecting people's depression dynamics?
    %\blue{It would be also good if we could sell the computational model here, for instance, how the model is effective, how it scales, what happens if we have less or more data? The reason is that this would also help tell reviewers and readers that the model used is effective and efficient and brings introduces some technical depth in the paper. We could separate the above questions as studying the dynamics as already done so far, and then studying the model effectiveness and efficiency.}
\end{itemize}

This paper aims to examine community depression dynamics due to COVID-19 pandemic in Australia. A new approach based on multi-modal features from tweets and term frequency-inverse document frequency (TF-IDF) is proposed to build a depression classification model. Multi-modal features aim to capture depression cues from emotion, topic and domain-specific perspectives. TF-IDF is simple, scalable, and has proven to be very effective to model a wide-range of lexical datasets including large and small datasets. In contrast, recent computationally demanding frameworks such as deep learning, which are usually generated rely upon large datasets because that helps give them faithful co-occurrence statistics might not be obtained from sparse and short texts from user-generated content such as tweets. Our approach uses the TF-IDF model can generalise well in various situations where both small and large datasets can be used leading to a reliable and scalable model. After building the model for depression classification, Twitter data in the state of New South Wales in Australia are then collected and input to our depression classification model to extract depression polarities which may be affected by COVID-19 and related events during the COVID-19 period. The contributions of this paper primarily include:

\begin{itemize}
    \item Novel multi-modal features including emotion, topic and domain-specific features are extracted to describe depression in a comprehensive approach;
    \item A faithful depression classification model based on TF-IDF, which is simple and scalable in generalisation, is proposed to detect depressions in short texts such as tweets;
    \item Instead of the depression examination of the whole country, a fine-grained analysis of depression in local government areas of a state in Australia is investigated;
    \item The links between the community depression and measures implemented by the government during the COVID-19 pandemic are examined.
\end{itemize}

To the best of our knowledge, this study is the first work to investigate the community depression dynamics across the COVID-19 pandemic, as well as to conduct the fine-grained analysis of depression and demonstrate the links of depression dynamics with measures implemented by the government and the COVID-19 itself during the pandemic. 

%\blue{tf.idf has existed in the Information Retrieval community for very long. It is very important to highlight why a simple tf.idf based approach was chosen as a modelling framework and why not other methods such as those based on deep learning which is getting increasingly popular these days. I think we could write that tf.idf is simple, scalable, and has proven to model a wide-range of datasets including large and small datasets. Recent computational frameworks such as deep learning are usually dependent on large datasets with faithful co-occurrence statistics which we might not get in sparse and short texts obtained from user-generated content such as twitter. So, our method could generalise well in various situations where we have both small and large datasets and thus is a faithful model. I also think that the way we could sell the paper could be different, for instance, focus more on multi-modality than tf.idf and say that tf.idf is a popular widely-used measure which we have adopted in this problem scenario. I also think we should mention the novelty of the paper in pointers as the research questions mentioned above. When we mention the novelty clearly, we then say that this is the first work.}

The remainder of the paper is organized as follows. We firstly review the related work in Section II and introduce the collected real-world dataset in Section III.
After that, we demonstrate our proposed method for COVID-19 depression analysis in section IV and present the experiments and verify the performance of the proposed model in Section V. The proposed novel model is then used to detect community depressions in New South Wales in Australia in Section VI. Finally, this work is concluded with an outlook on future work in Section VII.

\section{Related Work}\label{Relatedwork}

%\tempnotes{Shuiqiao works on this section.}

In this section, we review the related work for depression detection. We also highlight how our work differs from these existing approaches.

\subsection{Machine learning based depression detection}

Social media has long been used as the data source for depression detection due to the largely available user-generated text data \cite{sadeque2018measuring,kolliakou2020mental}. 
The shared text data and the social behaviour of the social network users are assumed to contain clues for identifying depressed users. To find the depression pattern for social media users, many works have been done to adopt traditional machine learning models such as Support Vector Machine (SVM) and J48 for depression classification and detection based on different feature engineering techniques. 
%\blue{this is what I have been referring to above, generalise it first and then say that we apply our model on Twitter data.}
%\blue{would be great if we could have examples here, a good presentation would be some sample tweet screenshots from twitter or Facebook with names anonymous.}.
For example, Wang et al. \cite{wang2013depression} have proposed a binary depression detection model for  Chinese Sina micro-blogs to classify the posts as depressed or non-depressed. Based on the features extracted from the content of the micro-blog such as the sentiment polarity of sub-sentences, the users' online interactions with others and the user behaviours, they have trained J48 tree, Bayes network, and rule-based decision table as classifiers for depression detection. 
%\blue{I see in the LaTex comments that there is a subsection created. Usually, if there are several works, it is better to break the related work into subheadings and topics and cluster them accordingly.} 

%\blue{Above work also seems closely related to the modelling perspective. It would be good to mention explicitly what is different in this work in comparison to our work? Mentioning the pros and cons of those works and compare them with the advantages of our method would be simply great. Remember that we have to sell this paper to the reviewers.}

De Choudhury \cite{de2013predicting} et al. have investigated to predict depression for social media users based on Twitter and found that social media contain meaningful indicators for predicting the onset of depressions among individual users. To get the ground truth of users' suffered depression history, De Choudhury et al. adopted the crowdsourcing to collect Twitter users who have been diagnosed with clinical (Major Depressive Disorder) MDD based on the CES-D2 (Center for Epidemiologic Studies Depression Scale) screening test. Then, to link the depression symptoms with the social media data, they extracted several measures such as user engagement and emotion, egocentric social graph, linguistic style, depressive language user, and the mentions of antidepressant medications from users' social media history for one year. Finally, they trained SVM as the depression classifier base on the ground truth and the extracted features for the tested Twitter users with prediction accuracy around 70\%.  Similarly, Tsugawa et al. \cite{tsugawa2015recognizing} have investigated in recognizing depression from the Twitter activities. To get the ground truth data of users' depression degree record, they chose the web-based questionnaire for facts collection. Then, similar features such as topic features extracted using topic modelling like Latent Dirichlet Allocation (LD), polarities of words and tweet frequency are extracted from the Twitter users' activity histories for training an SVM classifier. 

Later, Yang et al.  \cite{yang2015gis} have proposed to analyse the spatial patterns of depressed Twitter users based on Geographic Information Systems (GIS) technologies. They firstly adopted Non-negative Matrix Factorization (NMF) to identify the depressed tweets from online users. Then, they exploited the geo-tagged information as the indicator of users' geographical location and analyzed the spatial patterns of depressed users.
Shen et al. \cite{shen2017depression} have made efforts to explore comprehensive multi-modal features for depression detection. They adopted six groups of discriminant depression-oriented features extracted from users' social network, profile, visual content, tweets' emotions, tweets' topics and domain-specific knowledge are used as representations for Twitter users. Then, a dictionary learning model is used to learn from the features and capture the jointly cross-modality relatedness for a fused sparse representation and train a classifier to predict the depressed users.

Different from these work in feature extraction which may be either impractical like using crowdsourcing to manually label each user or too customized like extracting different kinds of social behaviours of users, we propose to combine the robust and simple text feature representation method: Term Frequency-Inverse Document Frequency (TF-IDF) with other multi-modal features such as topics and emotions to represent the text data.

\subsection{Deep learning based depression detection}

More recently, the rapidly developed deep learning techniques have also been used for depression detection. 
For example, Shen et al.  \cite{shen2018cross} have proposed a cross-domain deep neural network with feature transformation and combination strategy for transfer learning of depressive features from different domains. They have argued that the two major challenges regarding cross-domain learning are defined as isomerism and divergence and proposed DNN-FATC which includes Feature Normalization \& Alignment (FNA) and Divergent Feature Conversion (DFC) to better transfer learning. Orabi et al. \cite{orabi_deep_2018} have proposed to explore the word embedding techniques where they used pre-trained Skip-Gram (SG) and Continuous Bag-of-Words (CBOW) models currently implemented in the word2vec \cite{mikolov2013distributed} package for better textual feature extraction to train a neural network-based classifier.  We have also seen researchers using deep reinforcement learning to depression detection on Twitter. For instance, Gui et al. \cite{gui2019cooperative} have proposed a cooperative multi-agent model to jointly learn the textual and visual information for accurate depression inference. In their proposed method, the text feature is extracted using a Gated Recurrent Unit (GRU) and the visual feature is extracted using Convolutional Neural Networks (CNN). The selection for useful features from GRU and CNN is designed as policy gradient agents and trained by a centralized critic that implements difference rewards.

Even though deep learning has become the dominated method for many classifications or prediction tasks, it cannot guarantee that deep learning is feasible on any tasks. For example, \cite{nguyen-grishman-2015-relation} compared a bunch of methods including different conventional methods and deep learning methods for information extraction tasks from a text corpus. We are not surprised to find that deep learning did not outperform the conventional methods as expected. Furthermore, deep learning techniques are usually dependent on large datasets with faithful co-occurrence statistics which might not be obtained from sparse and short texts especially online user-generated content such as tweets which is characterised by relatively limited labelled text in the depression study. While TF-IDF is simple and scalable, it has proven to model a wide-range of datasets including large and small datasets. Therefore, in our study, we propose to adopt the traditional classification methods with TF-IDF for depression detection to obtain a faithful model with a strong generalisation ability in various situations of small and large datasets. Most importantly, even non-experts such as those from the government and NGOs could easily understand the intricacies of the model and apply our model on their data to detect community dynamics in their region and make further decisions accordingly.

%we are not surprised to find that the neural networks did outperform the conventional methods as much as expected. 
%Hence, in this work, 

\subsection{Depression detection due to COVID-19}

The impact of COVID-19 on people's mental health has been recently reported in various research. For instance, Galea et al. \cite{galea2020mental} have pointed out that the mental health consequences due to this pandemic are applicable in both the short and long term. They have appealed to develop ways to intervene with the inevitability of loneliness and its consequence as people are physically and socially isolated. Huang et al. \cite{huang2020generalized} have exploited a web-based cross-sectional survey based on the National Internet Survey on Emotional and Mental Health (NISEMH) to investigate people's mental health status in China by spreading the questionnaire on Wechat (a popular social media platform in China). To infer the depression symptoms for the anonymous participants, they have adopted a predefined Epidemiology Scale to identify whether participants had depressive symptoms. Similarly, Ni et al. \cite{ni2020mental} have conducted a survey on the online platform to investigate the mental health of 1577 community-based adults and 214 health professionals during the epidemic and lockdown. The results show that around one-fifth of respondents reported probable anxiety and probable depression. These works are mainly based on questionnaires and pre-defined mental heath scale models for inference. In contrast, our proposed work relies on detecting depression from social media data automatically which shows advantages in monitoring a large number of people's mental health states. 

\section{Data}
%\tempnotes{Jianlong and Shuiqiao work on this section.}

\subsection{Study location}

In this study, a case study for analysing depression dynamics during the COVID-19 pandemic in the state of New South Wales (NSW) in Australia is conducted. NSW has a large population of around 8.1 million people according to the census in September 2019 from Australian Bureau of Statistics\footnote{\url{https://www.abs.gov.au/}}. The NSW's capital city Sydney is Australia's most populated city with a population of over 5.3 million people. The Local Government Areas (LGAs) are the third tier of government in the Australian state (the three tiers are federal, state, and local government). 

\subsection{Data collection}
To analyse the dynamics of depression during the COVID-19 pandemic period at a fine-grained level, we collected tweets from Twitter users who live in different LGAs of NSW in Australia. The time span of the collected tweets is from 1 January 2020 to 22 May 2020 which covers dates that the first confirmed case of coronavirus was reported in NSW (25 January 2020) and the first time that the NSW premier announced the relaxing for the lockdown policy (10 Mary 2020). There are 128 LGAs in NSW. In this study, Twitter data were collected for each LGA separately so that the depression dynamics can be analysed and compared for LGAs. Twitter data were collected through the user timeline crawling API \textit{user\_timeline}\footnote{\url{http://docs.tweepy.org/en/latest/api.html#api-reference}} in Tweepy which is a python wrapper for the official Twitter API \footnote{\url{https://developer.twitter.com/en/docs/api-reference-index}}.  Table \ref{tab:lga_user_num} shows the summary of the collected tweet dataset. In summary, 94,707,264 tweets were collected with averagely 739,901 tweets for each LGA during the study period. Datasets of COVID-19 tests and confirmed cases in NSW during the study period were collected from DATA.NSW \footnote{\url{https://data.nsw.gov.au/}}. %\blue{Was the Twitter data collected using the Twitter API? If yes, I think it would be more comprehensive to mention that the API was used and how the API was used including any code used to connect to the API. The code link could be mentioned with code on Github.}

\begin{table}
   \caption{Summary of the collected Twitter dataset.}
    \centering
    \begin{tabular}{l|l}
    \hline
    \textbf{Description} &  \textbf{Size}\\
    \hline
    Total Twitter users &  183,104\\
         Average Twitter user per LGA & 1,430.5  \\ 
         Average tweets per LGA& 739,900.5\\
         Total tweets & 94,707,264\\
         \hline
    \end{tabular}
    \label{tab:lga_user_num}
\end{table}

\begin{figure*}
  \centering
  \includegraphics[width=0.99\linewidth]{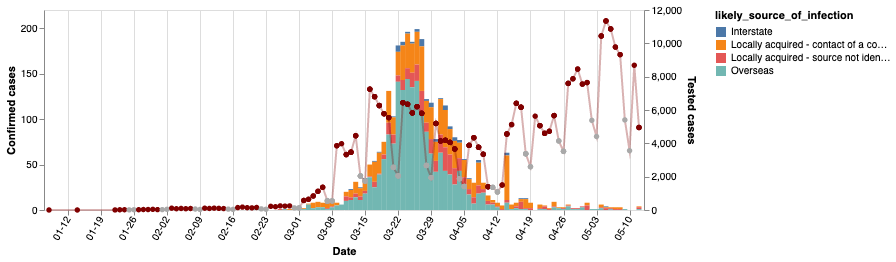}
  \caption{The tests and confirmed cases of COVID-19 in NSW until 22 May 2020.}
\label{fig:nsw_test_confirmed_cases}       
\end{figure*}

Fig. \ref{fig:nsw_test_confirmed_cases} shows the overview of the number of tests (polylines with dots) and confirmed cases (bars) of COVID-19 in NSW over the study period. It demonstrates that there usually had test peaks at the beginning of each week and had fewer numbers at the weekend, which well aligns with the people's living habits in Australia. It shows that the outbreak peak of COVID-19 in NSW was on 26 March 2020 and tests were significantly increased after 13 April 2020. It also shows that most of the confirmed cases were originally related to overseas.

\subsection{Dataset for depression model training}
In order to detect depression at the tweet level, we created two labelled datasets for both depressed and non-depressed tweets:
\begin{itemize}
    \item Positive tweets (depressed): we used a previous work dataset from \cite{shen2017depression}. Shen et al. \cite{shen2017depression} published around 300K tweets with 1400 depressed users. In order to have a better performance, we increased the number of positive tweets by crawling additional 600K tweets from the Twitter keyword streaming API. We adopted the same keywords search that selected by \cite{shen2017depression} where users identified themselves as depressed, and we also used a regular expression to find positive tweets (e.g. I'm depressed AND suicide). 
    \item Negative tweets (non-depressed): In order to balance the negative tweets with the positive tweets, we randomly selected 900K tweets which were not labelled as depressed from the collected tweets. Table \ref{tab:approach_data} shows the summary of labelled data used to train the depression model.
\end{itemize}

\begin{table}[!htb]
   \caption{Summary of labelled data used to train depression model.}
    \centering
    \begin{tabular}{l|l}
    \hline
    \textbf{Description} & \textbf{Size}\\
    \hline
        Depressed tweets  &  $\sim$ 900K\\
         Non-Depressed tweets & $\sim$ 900K \\ 
         \hline
    \end{tabular}
    \label{tab:approach_data}
\end{table}

After the collection of experimental data, features need to be extracted from social media text. However, because of the free-style nature of social media text, it is hard to extract useful features directly from raw text data and apply them to a classifier. The raw text data also affect the efficiency of extracting reliable features, and it makes it difficult to find word similarities and apply semantic analysis. Therefore, raw data must be pre-processed before feature extraction, in order to clean and filter data and ensure the data quality for feature extraction. Pre-processing may also involve other procedures such as text normalization. 

%Generally, the researchers normalize their text by applying several procedures.

Natural Language Processing (NLP) toolkit has been widely used for text pre-processing due to its high-quality processing capabilities such as processing sentimental analysis datasets \cite{horecki2015natural}. Natural Language Processing Toolkit (NLTK) library is considered as one of the most powerful NLP libraries in Python programming. NLTK contains packages that make data processing with human language easily and is used widely by various researchers for text data pre-processing. Therefore, before feeding our data to the model, we used NLTK to remove user mentions, URL links and punctuation from tweets. Furthermore, we removed common words from each tweet such as ``the'', ``an'', etc.). There are various reasons which have been mentioned in the literature where removing the stop words has had a positive impact on the model's quantitative performance, for instance, sometimes stop words deteriorate the classifications performance, sometimes they also have a huge impact on the model efficiency because these stop words increase the parameter space of the model, among various other reasons. NLTK has a set of stop words which enable removing them from any text data easily. Finally, we stem tweets using NLTK using the Porter Stemmer.

\section{Our Model}
%\subsection{Depression model}

\subsection{Proposed method}
\begin{figure*}[!htb]
 \centering
  \includegraphics[width=0.75\textwidth]{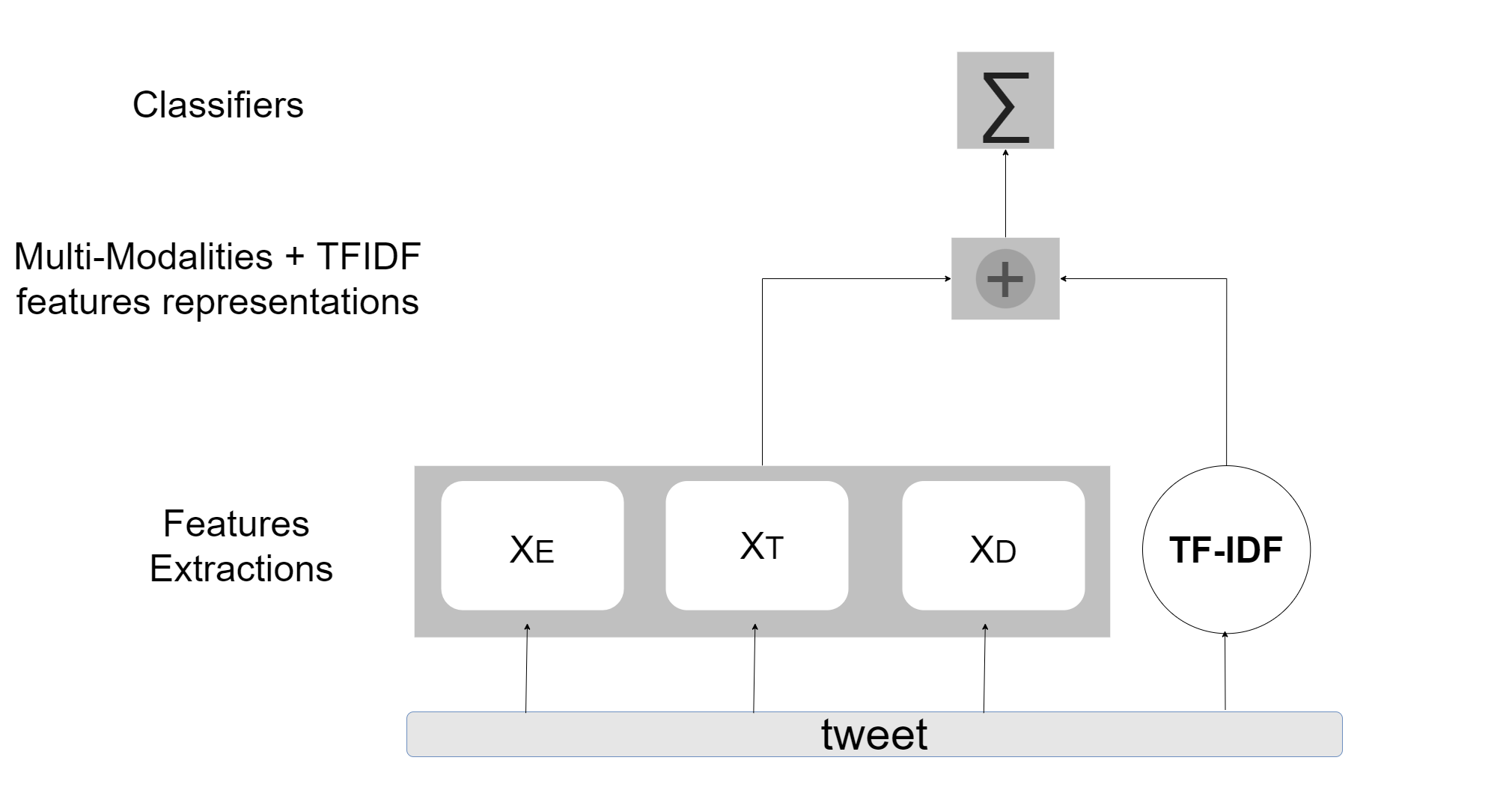}
  \caption{The proposed framework to detect depressed tweets during the COVID-19}
\label{fig:detection_approach}       
 \end{figure*}

In this study, two sets of features are extracted from raw text and used to represent tweet. Since extracting features would be challenging due to the short length of the tweets and single tweet does not provide sufficient word occurrences, we, therefore, combine multi-modal feature with Term Frequency-Inverse Document Frequency (TF-IDF) feature to analyze depressed tweets. Our proposed framework is shown in Fig. \ref{fig:detection_approach}.

% this is the figure of framework
%\input{framework}

\subsection{Multi-modal features}
User behaviours at the tweet level could be modelled by the linguistic features of tweets posted by the user. Inspired by \cite{shen2017depression}, we defined a set of features consisting of three modalities. These three modalities of features are as follows:

\begin{itemize}
\item \textbf{Emotional features}: The emotion of depressed people is usually different from non-depressed people, which influences their posts on social media. In this work, we studied user positive and negative emoji in each tweet to represent emotional features. Furthermore, users in social media often use a lot of slang and short words, which also convey positive and negative emotions \cite{novak2015sentiment}. In this study, positive and negative emotion features are also extracted based on those slang and short words.

\item \textbf{Topic-level features}: We adopted Latent Dirichlet Allocation (LDA) \cite{blei2003latent} to extract topic-level features since LDA is the most popular method used in the topic modelling to extract topics from text, which could be considered as a high-level latent structure of content. The idea of LDA is based on the assumption of a mixture of topics forms documents, each of which generates words based on their Dirichlet distribution of probability. Given the scope of the tweet content, we defined 25 latent topics in this study, which topic number is often adopted in other studies. We have also found that this number of topics gives satisfactory results in our experiments. we implemented LDA in Python with Scikit-Learn. %\blue{IMPORTANT: Which LDA implementation was used? This is important because different implementations might use different posterior inference algorithm.}
%\blue{Some reviewers could comment why did we use the LDA model here and why not Biterm topic model (BTM). If you have time if you could experiment with BTM and compare results with the LDA model. If there is no time, then you could write that we have used LDA because the model has been popularly used in latent topic modelling. To be on the safer side, it is better to include BTM.}

\item \textbf{Domain specific features}: Diagnostic and Statistical Manual of Mental Disorders 4th Edition (also called DSM-IV) is a manual published by the American Psychiatric Association (APA) that includes almost all currently recognized mental health disorder symptoms \cite{oconnor_screening_2009}. We, therefore, chose the DSM-IV Criteria for Major Depressive Disorder to describe keywords related to the nine depressive symptoms. Pre-trained word2vec (Gensim pre-trained model based on Wikipedia corpus) was used in this study to extend our keywords that are similar to these symptoms. We also extracted ``Antidepressant'' by creating a complete list of clinically approved prescription antidepressants in the world. %\tempnotes{Hamad, please add a citation on antidepressants here}. 
\begin{enumerate}
  \item Depressed mood.
  \item Loss of interest
  \item Weight or appetite change
  \item Sleep disturbance
  \item Psychomotor changes
  \item Fatigue or loss of energy
  \item Feel Worthlessness
  \item Reduced concentration
  \item Suicidal ideation
\end{enumerate}

\end{itemize}
For a given sample of tweet, the multi-modal features are represented as \(X_{t1}, X_{t2}, X_{t3},\ldots, X_{tn}\), where  \(\mathit X_{ti} \in \,\, {\mathbb R}^{\mathbf d}\) is the \(d\)-dimensional feature for the \(i\)-th modality for each tweet and $n$ is the size of the combined feature space, which is 21 in this study. %\tempnotes{Hamad, please confirm this description is right} . 

% After extracting multi-modality features, the features will be represented, give a sample of tweet tn the multi-modal features representation. Given the original feature representation for each tweet \(X_{t1}, X_{t2}, X_{t3},\ldots,X_{tn}\), where  \(\mathit X_{tn}\in \,\,\mathbb {R}^{\mathbf d}\) is the d-dimensional features for the n-th modal for each tweet and n is the length of all the features with the total of 21.

\subsection{TF-IDF}
We first review the definitions of Term Frequency and Inverse Document Frequency below.

\textbf{Definition 1. } \textbf{Term Frequency (TF)}: consider the tweets in Bag-of-Word model (BoW), where each tweet is modeled as a sequence of words without order info. Apparently in BoW scheme, a word in a tweet with occurrence frequency of 10 is more important than a term with that frequency of 1, but not proportional to the frequency. Here we use the log term frequency $ltf$ as defined by:
% \begin{center}
% log term frequency (t,d) defined as:
% \end{center} 
\begin{equation}
\label{equ:tf}
ltf_{(t,d)} = 1+log(1+tf_{(t,d)})
\end{equation}

\noindent where $tf_{(t,d)}$ represents occurrence number of the term $t$ in a tweet $d$. 

%They were able to achieve high precision and accuracy with five different Machine learning algorithms.

\textbf{Definition 2. } \textbf{Inverse Document Frequency (IDF)}: It uses the frequency of the term in the whole collection for weighting the significance of the term in light of inverse effect. Therefore under this scheme, the IDF value of a rare word is higher, whereas lower for a frequent term, i.e. weighting more on the distinct words.
The log IDF which measures the informativeness of a term is defined as:
\begin{equation}
\label{equ:idf}
idf_{t}=log_{10} \frac{N}{df_{t}}
\end{equation}

\noindent where $N$ represents the total number of tweets in the tweet corpus, and $df_{t}$ the number of tweets containing the term $t$.

%\blue{the - sign is confusing. Can we use . instead of - everywhere?}
Therefore, TF-IDF is calculated by combining TF and IDF as represented in Eq.\ref{equ:tf-idf}. 
\begin{equation}
\label{equ:tf-idf}
tfidf_{t}=tf_{t} * idf_{t}
\end{equation}

In order to extract relevant information from each tweet and reduce the amount of redundant data without losing important information, we combined multi-modalities with TF-IDF.
TF-IDF is a numerical statistic metric to reflect the importance of a word in a list or corpus, which is widely studied in relevant work. Choudhury et al. \cite{de2013predicting} applied TF-IDF to words in Wikipedia to remove extremely frequent terms and then used the top words with high TF-IDF. The approach helped them to assess the frequency of uses of depression terms that appear on each user's Twitter posts on a given day. In \cite{singh2019framework} the authors have used TF-IDF in their model to compare the difference of performance by feeding TF-IDF into five machine learning models. It was found that all of them can achieve very good performance. However, one weakness of BoW is unable to consider the word position in the text, semantics, co-occurrences in different documents. Therefore, TF-IDF is only useful as a lexical level feature.

%We can see that TF-IDF is easy to compute besides it can extract the most descriptive terms in a document. It can also be used to compute the similarity between two documents. 
\subsection{Modeling depression in tweets}
%\subsubsection{Experimental Setup} 
Two labelled datasets as introduced in the previous section are used to train the depression classification model for tweets. Here we only use English tweets to train the model, and all non-English tweets are excluded. We also exclude any tweets with a length shorter than five words since these tweets could only introduce noise and influence the effectiveness of the model negatively. Three mainstream classification methods are used in this study to compare their performance, namely Logistic Regression (LR), Linear Discriminant Analysis (LDA), and Gaussian Naive Bayes (GNB) .We used scikit-learn libraries to import the three classification methods. The classification performance by these three methods were evaluated by 5-fold cross-validation. The experiments are conducted using Python 3.6.3 with 16-cores CPU.

%\blue{IMPORTANT: What is missing in this paper are experimental settings. How many iterations each model were executed? What were the hyperparameter settings? Which implementation did we use? Where are the codes, and if we have implemented our own code, is it publicly available? These questions can only be answered Hamad. It is much safer if we share the codes, because parameter settings would be mentioned there. If we have used an existing implementation, we simply mention what settings we used in that existing implementation.}

%\tempnotes{Hamad, from the result statement, you only use LR, LDA, GNB for model training, there is no deep learning approaches. Why Tensorflow is mentioned here. If it is used, could you please also indicate more details on the use of Tensorflow, such as settings of tensorflow?} \blue{I think we would also need to a justification why these models have only been chosen? What advantages they have for instance when compared to SVM? Also, we might need to justify why neural networks were not chosen? There are some nice codes which could help with neural networks experiments: https://github.com/nadbordrozd/text-top-model.}

%\subsubsection{Metrics}
We evaluate the classification models by using measure of Accuracy (ACC.), Recall (Rec.), Macro-averaged Precision (Prec.), and Macro-averaged F1-Measure (F1).

\section{Classification Evaluation Results}
We firstly evaluate how well the existing models can detect depressed tweets. After extracting the feature representations of each tweet, three different classification models are trained with the labelled data. We adopted a ratio of 75:25 to split the labelled data into training and testing data. We performed experiments by using TF-IDF, multi-modality, and combined features under three different classification models. 

We compare the performance of models using only Multi-Modalities (MM) features with the three different classifiers as shown in Table \ref{tab:modal}. We found that Gaussian Naive Bayes for MM features obtained the highest Precision score compared to other classifiers, and Logistic Regression performs better than the other two classifiers in terms of Recall, F1-Score, and Accuracy.

\begin{table}
   \caption{The performance of tweet depression detection based on multi-modalties only.}
    \centering
    \begin{tabular}{|c|c|c|r|r|r|r|r|}
    \hline
    \textbf{Features} &
    \textbf{Method} &
    \textbf{Precision} &
    \textbf{Recall} &
    \textbf{F1-Score} &
    \textbf{Accuracy}\\
    \hline
     &  LR & 
     0.842 & 0.828 & 0.832 & 0.833\\
         Multi-Modal & LDA &
    0.843 & 0.816 & 0.820 & 0.824\\ 
           & GNB &
    0.873 & 0.814 & 0.818 & 0.825\\
    %  &  LR & 
    %  0.84 & 0.83 & 0.83 & 0.833007\\
    %      Multi-Modal. & LDA &
    % 0.84 & 0.82 & 0.82 & 0.825119\\ 
    %       & GNB &
    % 0.87 & 0.81 & 0.81 & 0.825088\\
         \hline
    \end{tabular}
    \label{tab:modal}
\end{table}

Furthermore, in order to see how different features impact the classification performance, we use the TF-IDF standalone with three classifiers for comparison and the results are as shown in Table \ref{tab:tf}. It shows that all three classifiers can achieve satisfactory classification results. LR and LDA shared the highest Precision and F1-Score. LR competes with the other two classifiers according to the Recall and Accuracy.

\begin{table}
   \caption{The performance of tweet depression detection based on TF-IDF only.%\blue{Values in the last column need to rounded off to two decimal places for consistency with rest of the numbers. It would be good if all results are in three decimal places.}
   }
    \centering
    \begin{tabular}{|c|c|c|r|r|r|r|r|}
    \hline
    \textbf{Features} &
    \textbf{Method} &
    \textbf{Precision} &
    \textbf{Recall} &
    \textbf{F1-Score} &
    \textbf{Accuracy}\\
    \hline
         &  LR & 
     0.908 & 0.896 & 0.900 & 0.901\\
         TF-IDF & LDA &
    0.906 & 0.893 & 0.897 & 0.898\\ 
           & GNB &
    0.891 & 0.873 & 0.877 & 0.879\\
    %  &  LR & 
    %  0.91 & 0.90 & 0.90 & 0.901022\\
    %      TF-IDF & LDA &
    % 0.91 & 0.89 & 0.90 & 0.898124\\ 
    %       & GNB &
    % 0.89 & 0.87 & 0.88 & 0.879432\\
         \hline
    \end{tabular}
    \label{tab:tf}
\end{table}

Table \ref{tab:tf_modal} shows the model performance when we concatenate both MM and TF-IDF features, and we can see that the model has improved the performance further slightly. One conclusion we can draw here is that TF-IDF textual feature can make main satisfactory contribution to detect depression tweets, while other modality can provide additional support. This could be attributed to the lexical importance to depressed-related tweets. Another reason why the overall combination of multi-modal features did not give us a big lift in the results could be because, as mentioned earlier, tweets are highly sparse, poorly phrased, short texts, and noisy content. Therefore, deriving semantic content from tweets, for instance, using the LDA model would always be very challenging to get a huge boost in the results. This is mainly because any statistical model relies on the co-occurrence statistics which might be poor in our case. However, we still see an improvement in the overall recall result, which is important in this case, because we are noticing a reduction in the number of false negative detection. This is mainly because of the interplay between all three features which suggest that these features are important and cannot be ignored.

\begin{table}
   \caption{The performance of tweet depression detection based on Multi-Modalties + TF-IDF.}
    \centering
    \begin{tabular}{|c|c|c|r|r|r|r|r|}
    \hline
    \textbf{Features } &
    \textbf{Method} &
    \textbf{Precision} &
    \textbf{Recall} &
    \textbf{F1-Score} &
    \textbf{Accuracy}\\
    \hline
     &  LR  & 
     0.908 & 0.899 & 0.902 & 0.903\\
         MM+TF-IDF & LDA &
    0.912 & 0.899 & 0.903 & 0.904\\ 
           & GNB &
    0.891 & 0.874 & 0.878 & 0.879\\
    %  &  LR  & 
    %  0.91 & 0.90 & 0.90 & 0.903191\\
    %      MM+TF-IDF & LDA &
    % 0.91 & 0.90 & 0.90 & 0.903467\\ 
    %       & GNB &
    % 0.89 & 0.88 & 0.88 & 0.880558\\
         \hline
    \end{tabular}
    \label{tab:tf_modal}
\end{table}

% The figure in the next section is put here to make it appear on the same page with its related text
\begin{figure*}[!tb]
  \centering
  \includegraphics[width=0.99\linewidth]{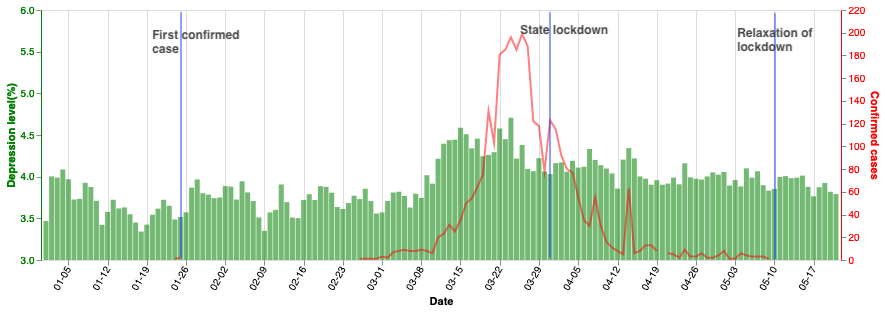}
  \caption{The community depression dynamics in NSW between 1 January 2020 and 22 May 2020.}
\label{fig:nsw_depression}       
\end{figure*}

\section{Detecting Depression due to COVID-19}

After having testified our classification model, we utilise our approach to detect depressed tweets from different LGAs of NSW, Australia. Since our model deals only with English tweets, we had to exclude tweets in all other languages and input English tweets only into our model to predict depression. We ended up with 49 million tweets from 128 LGAs in NSW. We fed the LR with (MM + TF-IDF) model with these tweets, and the model found that nearly 2 million tweets were classified as depressed tweets. In this section, we show the depression dynamics in NSW during the study period between 1 January 2020 and 22 May 2020. The depression dynamics in different LGAs in NSW are also analysed to demonstrate how COVID-19 pandemic may affect people's mental health.

\subsection{Depression dynamics in NSW}

Fig.~\ref{fig:nsw_depression} presents the overall community depression dynamics in NSW with the confirmed cases of COVID-19 together during the study period between 1 January 2020 and 22 May 2020. ``Depression level'' refers to the proportion of the number of depressed tweets over the whole number of tweets each day. From this figure, we can find that people showed a low level of depression during the period before the significant increase in the confirmed cases of COVID-19 until 8 March 2020 in NSW. People's depression level was significantly increased with a significant increase in confirmed cases of COVID-19. The depression level reached to the peak during the peak outbreak period of COVID-19 on 26 March 2020. After that, people's depression level decreased significantly for a short period and then kept relatively stable with some short fluctuations. Overall, the analysis clearly shows that people became more depressed after the outbreak of COVID-19 on 8 March 2020 in NSW.

When we drill down into details of Fig.~\ref{fig:nsw_depression}, it was found that people's depression was much sensitive to sharp changes of confirmed cases of COVID-19 on that day or after. For example, people's depression level had sharp changes around days of 10 March, 25 March, 30 March, and 14 April 2020 (there were sharp changes of confirmed cases of COVID-19 on these days). The sharp increase of confirmed cases of COVID-19 usually resulted in the sharp increase of depression levels (i.e. people became more depressed because of the sharp increase of confirmed cases of COVID-19).

% this figure in the next subsection is put here in order to make it appear on the top of the page
\begin{figure*}[!htb]
  \centering
  \includegraphics[width=0.49\linewidth]{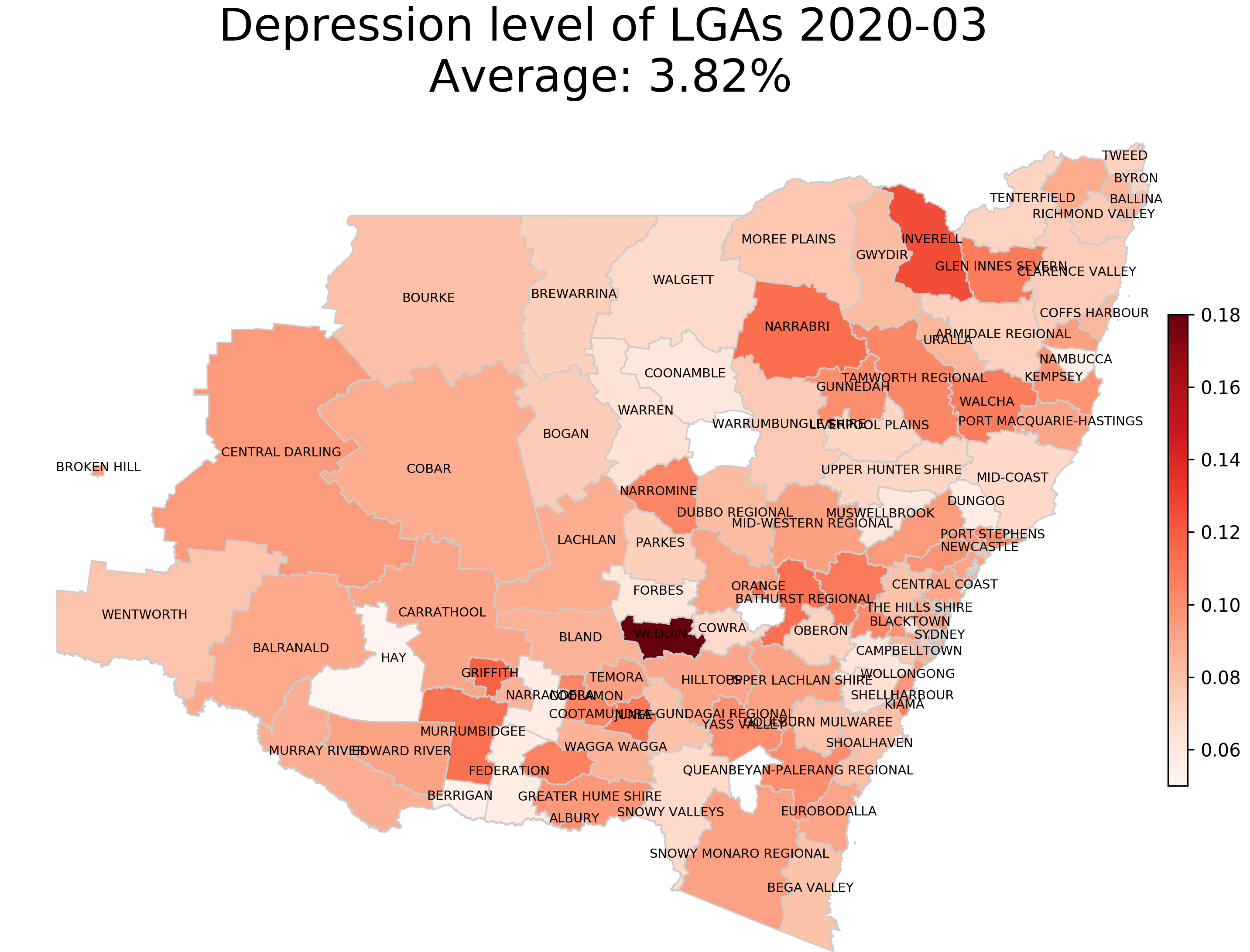}
  \includegraphics[width=0.49\linewidth]{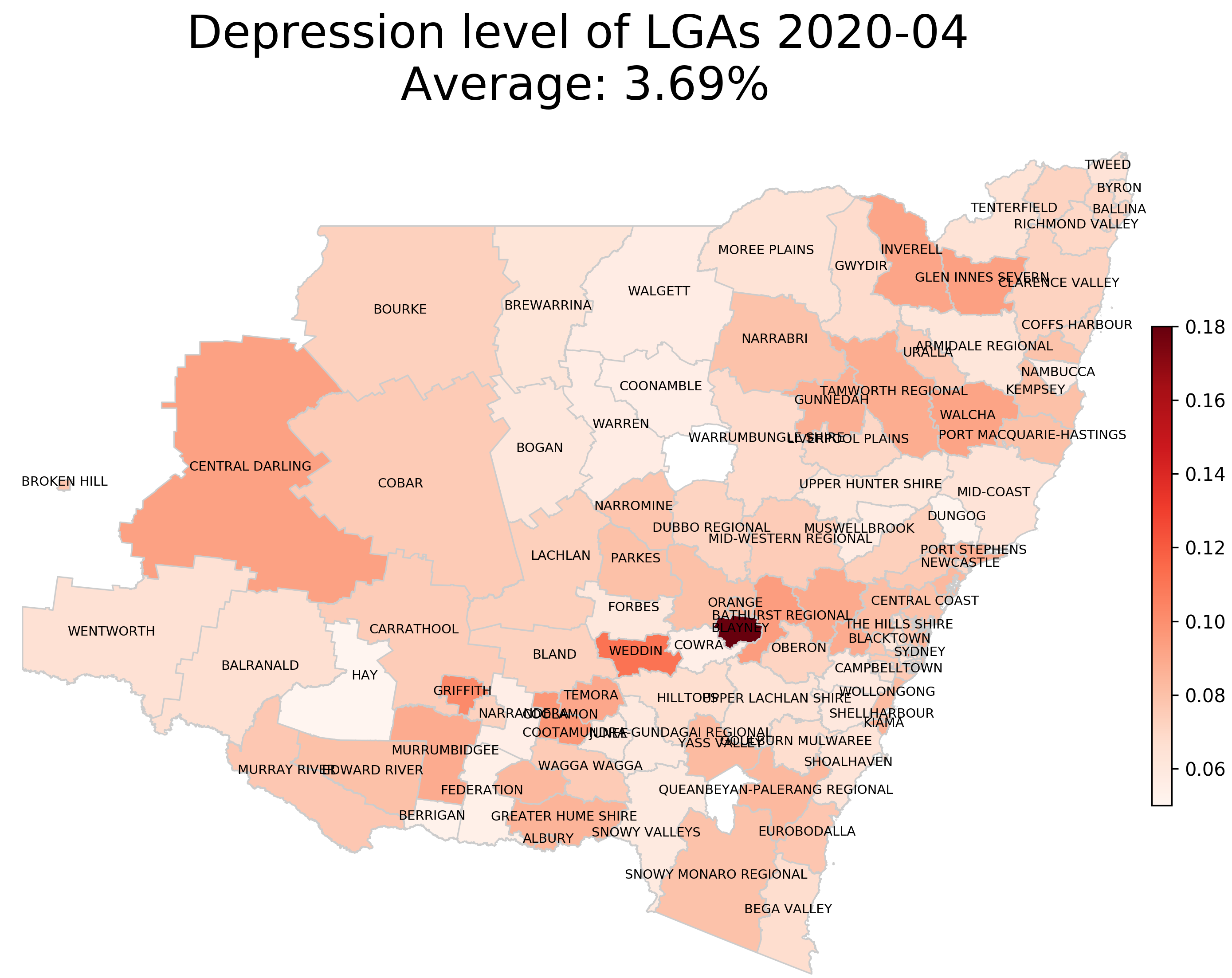}
  \caption{The choropleth maps of community depression in LGAs in NSW in March 2020 (left) and April 2020 (right).}
\label{fig:lga_depression_all} 
\end{figure*}

\subsection{Depression under implemented government measures and big events}

This subsection investigates the links between people's depression and implemented government measures for COVID-19 (such as lockdown) as well as big events during the pandemic.

By investigating the labelled topics with the hashtag in the collected Twitter data, it was found that the topics of ``lockdown'' and ``social-distancing'' were started to be discussed actively from 9 March when the government encouraged people to increase social-distancing in the life, and the NSW government officially announced the state lockdown on 30 March and the restrictions were begun from 31 March\footnote{\url{https://gazette.legislation.nsw.gov.au/so/download.w3p?id=Gazette_2020_2020-65.pdf}}. The NSW government announced the ease of restrictions on 10 May\footnote{\url{https://www.nsw.gov.au/media-releases/nsw-to-ease-restrictions-week-0}}. When we link these dates with depression levels as shown in Fig.~\ref{fig:nsw_depression}, it was found that people felt significantly more depressed when they started to actively discuss the lockdown restriction on 9 March. People were slightly more depressed after the official implementation of the state lockdown in NSW. The results revealed that the lockdown measure may make people more depressed. However, people still became more depressed even if after the relaxation of lockdown. This is maybe because people still worried about the spread of this severe virus due to the increased community activities.

We are also interested in whether people's depression was affected by big events during the COVID-19 period. For example, the Ruby Princess Cruise docked in Sydney Harbour on 18 March 2020. About 2700 passengers were allowed to disembark on 19 March without isolation or other measures although some passengers had COVID-19 symptoms at that time, which was considered to create a coronavirus hotbed in Australia. The Ruby Princess Cruise has been reported to be linked to at least 662 confirmed cases and 22 deaths of COVID-19\footnote{\label{fn:au_covid_time}\url{https://www.theguardian.com/world/2020/may/02/australias-coronavirus-lockdown-the-first-50-days}}.

We link the depression dynamics as shown in Fig.~\ref{fig:nsw_depression} with the important dates of the Ruby Princess Cruise (e.g. docking date, disembarking date) and actual timeline of the public reporting of confirmed cases and deaths as well as other events (e.g. police in NSW announced a criminal investigation into the Ruby Princess Cruise debacle on 5 April) related to the Ruby Princess \textsuperscript{\ref{fn:au_covid_time}}. We have not found significant changes in people's depression on those dates. This may imply that the big events did not cause people's depression changes significantly.

\subsection{Depression dynamics in LGAs}

We further analysed community depression dynamics in LGAs in NSW. Fig.~\ref{fig:lga_depression_all} shows examples of choropleth maps of community depression dynamics in LGAs in NSW for two different months which are March 2020 and April 2020. We can observe from the maps that the community depression level was different across different LGAs in each month. Furthermore, the community depression of each LGA changed in different months. On average, people in LGAs were more depressed in March than in April. This is maybe because the number of daily confirmed cases of COVID-19 was significantly increased to a peak and it was gradually decreased in April.     

\begin{figure*}[!htb]
  \centering
  \includegraphics[width=0.95\linewidth]{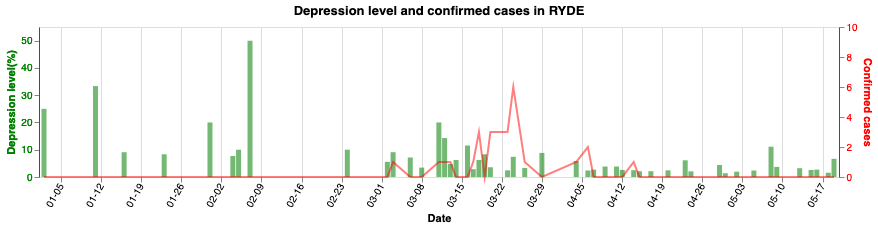}
  \includegraphics[width=0.95\linewidth]{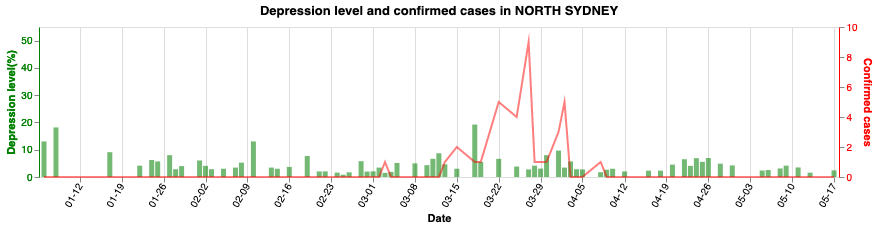}
  \includegraphics[width=0.95\linewidth]{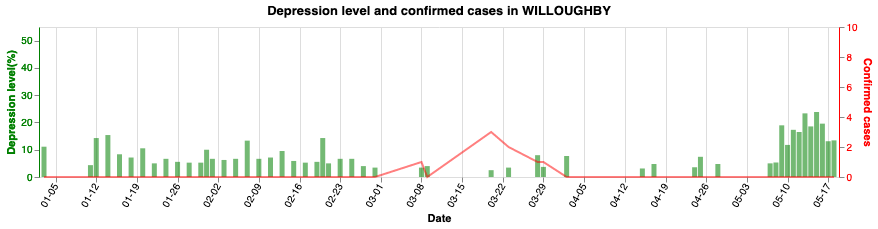}
  \caption{The community depression dynamics in Ryde, North Sydney, and Willoughby in Northern Sydney between 1 January 2020 and 22 May 2020.}
\label{fig:lga_depression_selected}       
\end{figure*}

We also dig into more details of depression changes of LGAs around Sydney City areas. For example, Ryde, North Sydney, and Willoughby are three neighbouring LGAs in Northern Sydney. Their community depression dynamics and corresponding confirmed cases of COVID-19 are shown in Fig.~\ref{fig:lga_depression_selected}, respectively. When comparing the dynamics in this figure, we can see that different LGAs showed different depression dynamics maybe because some events specifically related to that LGA. For example, in an aged care centre in Ryde LGA, a nurse and an 82-year-old elderly resident were first tested positive for coronavirus at the beginning of March\footnote{\url{https://www.smh.com.au/national/woman-catches-coronavirus-in-australia-40-sydney-hospital-staff-quarantined-20200304-p546lf.html}}. After that, a number of elderly residents in this aged care centre were tested positive for COVID-19 or even died. At the same time, a childcare centre and a hospital in this LGA have been reported positive COVID-19 cases in March. Many staff from the childcare centre and the hospital were asked to test the virus and conduct home isolation for 14 days. All these may result in the significant community depression changes in March 2020 as shown in Fig.~\ref{fig:lga_depression_selected}. For example, the depression level in Ryde LGA was changed significantly to a very high level on 10 March 2020 and 16 March 2020 respectively.

However, it was not found that the community depression dynamics in an LGA showed close relations with the dynamics of confirmed cases of COVID-19 in that LGA as we found in the state level. Maybe this is because the community depression dynamics in an LGA was affected largely by the confirmed cases in the overall state but not the local government area. This aligns with our common sense during the COVID-19 pandemic: even if our family is currently safe from COVID-19, we are still worrying about the life because of the continuing significant increases of COVID-19 all over the world especially in big countries.

% NSW Local Health Districts (LHD)
% Ryde is one of LGAs in the Northern Sydney LHD 
% \begin{figure}[!htb]
%   \centering
%   \includegraphics[width=0.85\linewidth]{northern_sydney_lhd.png}
%   \caption{The Northern Sydney LHD map\protect\footnotemark.}
% \label{fig:northern_sydney_lhd_map}       
% \end{figure}
% \footnotetext{\url{https://www.health.nsw.gov.au/lhd/Pages/lhd-maps.aspx}}

\subsection{Discussion}

COVID-19 pandemic has affected people's lives all over the world. Due to social-distancing measures and other restrictions implemented by the government, people often use social media such as Twitter for socialising. 

%This paper proposed a TF-IDF based approach to extract multi-modality features from user tweets, which are used to build classification models capable of capturing depression among users. The novel classification model was used to detect depression dynamics of people in NSW. 

The results of this study showed that our novel depression classification model can successfully detect community depression dynamics in NSW state level during the COVID-19 pandemic. It was found that people became more depressed after the outbreak of COVID-19 in NSW. People's depression level was much sensitive to sharp changes in confirmed cases of COVID-19, and the sharp increase of confirmed cases made people more depressed. When we conducted a fine-grained analysis of depression dynamics in LGAs in NSW, our novel model can also detect the differences of people's depression in LGAs as well as depression changes in each LGA at a different period. 

The study found that the policies/measures implemented by the government such as the state lockdown has had obvious impact on community depression. The implementation of the state lockdown made people more depressed; however, the relaxation of restriction also made people more depressed. This could be primarily because people are still worried about the spread of COVID-19 due to the increased community activities after the relaxation of restrictions. 

This study did not find the significant effects of big events such as the Ruby Princess Cruise ship coronavirus disaster in Sydney on the community depression. This is maybe because the data related to the disaster is small and passengers were also from other Australian states and even overseas besides NSW.

%\tempnotes{TO All: any discussions on the methods and others?}

%\blue{This is the comments for the research questions in introduction section from Shoaib. I also moved here for possible discussions on the model (JZHOU): It would be also good if we could sell the computational model here, for instance, how the model is effective, how it scales, what happens if we have less or more data? The reason is that this would also help tell reviewers and readers that the model used is effective and efficient and brings introduces some technical depth in the paper. We could separate the above questions as studying the dynamics as already done so far, and then studying the model effectiveness and efficiency.}

\section{Conclusion and Future Work}
%\tempnotes{Jianlong mainly works on this section.}
%The COVID-19 pandemic has affected people's lives all over the world. 
This paper conducted a comprehensive examination of the community depression dynamics in the state of NSW in Australia due to the COVID-19 pandemic. A novel depression classification model based on multi-modal features and TF-IDF was proposed to detect depression polarities from the Twitter text. By using Twitter data collected from each LGAs of NSW from 1 January 2020 until 22 May 2020 as input to our novel model, this paper investigated the fine-grained analysis of community depression dynamics in NSW. The results showed that people became more depressed after the outbreak of the COVID-19 pandemic. People's depression was also affected by the sharp changes in confirmed cases of COVID-19. Our model successfully detected depression dynamics because of the implementations of measures by the government. When we drilled down into LGAs, it was found that different LGAs showed different depression polarities during the timeframe of the tweets used in our study, and each LGA may have different depression polarity on different days. It was observed that the big health emergencies in an LGA had a significant impact on people's depression. However, we did not find significant effects of the confirmed cases of COVID-19 in an LGA on people's depressions in that LGA as we observed in the state level. These findings could help authorities such as governmental departments to manage community mental health disorders more objectively. The proposed approach can also help government authorities to learn the effectiveness of policies implemented.

In this special period of the COVID-19 pandemic, we focused on the effects of COVID-19 on people's depression dynamics. However, 
other factors such as unemployment, poverty, family relationship, personal health, and various others may also lead people to be depressed. Our future work will investigate how these factors may affect community depression dynamics. Furthermore, community depression will also be investigated using the topics over time model and using the temporal topics as multi-modal features. 
More recent and advanced classification models will be investigated to classify people's depression polarities.

% \section*{Acknowledgment}

\ifCLASSOPTIONcaptionsoff
  \newpage
\fi


\bibliographystyle{IEEEtran}
%\bibliography{IEEEabrv,sentiment,depression}
\bibliography{depression}
\end{document}